\documentclass{JHEP}

\input epsf

\usepackage{epsfig}

\usepackage{amssymb}
\usepackage{axodraw}

\newcommand{\bea}{\begin{eqnarray}}

\newcommand{\eea}{\end{eqnarray}}

\newcommand{\bean}{\begin{eqnarray*}}

\newcommand{\eean}{\end{eqnarray*}}

\newcommand{\be}{\begin{equation}}
\newcommand{\ee}{\end{equation}}
 
\newcommand{\beq}{\begin{equation}}
\newcommand{\eeq}{\end{equation}}
\newcommand{\bqa}{\begin{eqnarray}}
\newcommand{\eqa}{\end{eqnarray}}
\newcommand{\nl}{\nonumber \\}
\def\db#1{\bar D_{#1}}
\def\d#1{D_{#1}}
\def\tld#1{\tilde {#1}}

\def\eqn#1{Eq.~(\ref{#1})}

\def\Label#1{\label{#1}
 \smash{\hbox to0pt{\raise1ex\hbox{\tiny[#1]}\hss}}}

\preprint{CERN-PH-TH/2008-057}

\title{Optimizing the Reduction of One-Loop Amplitudes}

 \author{ P. Mastrolia${}^a$, G. Ossola${}^b$, C. G. Papadopoulos${}^b$, and R. Pittau${}^{c}$ \\

${}^a$ Theory Division, CERN, CH-1211 Geneva 23, Switzerland \\

${}^b$ Institute of Nuclear Physics, NCSR Demokritos, 15310 Athens, Greece\\

${}^c$ Departamento de F\'{i}sica Te\'orica y del Cosmos, CAPFE,
       Universidad de Granada, E-18071 Granada, Spain \\

~~~~~~~\\

 \email{  \hskip0.5cm} }

\abstract{
We present an optimization of the
reduction algorithm of one-loop amplitudes in terms of
master integrals. It is based on the exploitation
of the polynomial structure of the integrand
when evaluated at values of the loop-momentum
fulfilling multiple cut-conditions, as
emerged in the OPP-method.
The reconstruction of the polynomials,
needed for the complete reduction, is
rended very versatile by using a projection-technique
based on the Discrete Fourier Transform.
The novel implementation is applied
in the context of the NLO QCD corrections to $u \bar{d} \to W^+ W^- W^+$.
}

\begin{document}


\section{Introduction}
In the last few years, we have been witnessing a boost in developing
new ideas aiming to the efficient computation of one-loop amplitudes~\cite{Review},
as extensively reported in~\cite{LesHouches2007}.
Besides standard techniques, where the tensor reduction is explicitly performed,
new numerical and analytical developments, originally
inspired by unitarity arguments~\cite{Bern:1994zx, Bern:1994cg} have emerged.
The common features of the so called unitarity-based methods
\cite{Cachazo:2004by}
-\cite{Giele:2008ve}, 
is the change of perspective they propose: instead of focusing on 
the actual evaluation of complete integrals,
they pursue the determination of the {\em coefficients}
of the scalar one-loop functions contributing to the unknown integrals.
This possibility relies on the fact that the
basis of scalar function to express any one-loop integrals is known in terms of Boxes, Triangles,
Bubbles and (in massive theories) Tadpoles~\cite{MasterIntegrals}.
Schematically, one can write
a Master Equation for any one-loop amplitude ${\cal M}$ such as:
\bqa
\label{eq:MasterEqn}
{\cal M}=
\sum_i d_i {\rm ~Box}_i
+\sum_i c_i {\rm ~Triangle}_i
+\sum_i b_i {\rm ~Bubble}_i
+\sum_i a_i {\rm ~Tadpole}_i
\,,
\eqa
where $d_i$, $c_i$, $b_i$ and $a_i$ are the coefficients
to be determined. \\
Very recently in \cite{Britto:2008sw,Britto:2008vq}, by exploiting the properties of spinor-integration
of double-cuts of dimensionally regulated integrals
~\cite{Britto:2005ha,Britto:2006sj,Anastasiou:2006jv,Mastrolia:2006ki,Britto:2006fc,Anastasiou:2006gt,Britto:2006fc,Britto:2007tt}, general analytic formulas for the coefficients 
$d_i$, $c_i$, $b_i$ were presented. These formulas can
be evaluated at the occurrence, without performing any integration,
by specializing the value of input variables that are specific to the 
initial cut-integrand, which is
assembled from tree-level amplitudes.

Alternatively to any phase-space integration, in \cite{Ossola:2006us,Ossola:2007bb} it was proposed
a very efficient method for the reconstruction of the so-called 4-dimensional cut-constructible term of 
any scattering amplitude, corresponding to the poly-logarithmic structure arising when Eq.(\ref{eq:MasterEqn})
is expanded around 4-dimensions.
This method, by-now known as {\tt OPP}-reduction, allows the numerical reconstruction of (the 4-dimensional limit of) the
coefficients, $d_i$, $c_i$, $b_i$ and $a_i$, by solving a system of algebraic equations that are obtained by: {\it i)} the numerical evaluation of the {\it integrand} at explicit values of the
loop-variable, on the one side; 
{\it ii)} and the knowledge of the most general {\it polynomial} structure of the {\it integrand}
itself~\cite{intlevel}, on the other one. 
We remark that the values of the loop momentum used for the numerical evaluation of the integrand
are chosen among the set of solutions of the multiple-cut conditions, namely the solutions of the system of
equations obtained by imposing the vanishing of the denominators on each 4-dimensional cut. 

For the complete evaluation of scattering amplitudes, one has to consider that 
the 4-dimensional expansion of Eq.(\ref{eq:MasterEqn}) generates not only
a poly-logarithmic term, but as well a {\it rational term} which cannot
be detected by (massless) cuts in 4-dimensions. 
In \cite{Ossola:2008xq}, it has been recently shown that there are two sources of the rational terms: 
the first contribution, that is quite simple to calculate, originates from the generic $(n-4)$-dimensional structure of the numerator of any one-loop amplitude and it can be derived by using
appropriate Feynman rules within a tree-like computation. The second contribution
originates instead from the reduction of the $4-$dimensional part of the numerator in terms of the $n-$dimensional denominators appearing in the scalar integrals. This part, that is more subtle to 
extract, can be computed within the {\tt OPP}-method in a completely automatized way, following one of the approaches discussed in \cite{Ossola:2008xq} and numerically implemented in the public code {\tt CutTools} \cite{Ossola:2007ax} .\\
Alternatively,
the reconstruction of the rational term can be achieved as well by using
techniques like
direct computation~\cite{directcomp1,directcomp2}, 
by the bootstrapping method~\cite{bootstrap}, 
by cuts in $n$-dimensions~\cite{Britto:2008sw,Britto:2008vq}, or
by explicitly computing the
amplitude at different integer value of the space-time dimensions \cite{Giele:2008ve}.
In particular, in \cite{Ellis:2007br,Giele:2008ve} it has been proposed an extension of the {\tt OPP}-reduction,
implementing an integrand decomposition valid in $n$-dimension, rather than in 4-dimension, which
exposed a richer, yet {\it polynomial}, structure of the cut-integrand.

The efficiency of the {\tt OPP}-reduction has been shown in non-trivial applications, like
the 6-photon amplitudes with massless and massive fermion-loop \cite{Ossola:2007bb}, 
the virtual QCD correction to $q \bar{q} \to ZZZ$ \cite{LesHouches2007}, and the complete cross section
for the production of three vector bosons at LHC \cite{3bosons}.

\bigskip 

Within the {\tt OPP}-reduction, the coefficients of the master integrals can be simply
extracted by solving a system of numerical equations, 
rather than computing phase-space integrals.
Cut-by-cut, in a {\it top-down} algorithm, 
from quadruple- to single-cut, one can establish a system for extracting the coefficient
of each master-integral identified by the product of the cut-propagators.
The general structure of the integrand~\cite{intlevel} determines
the polynomial shape of the equations forming such a system.
In fact, by decomposing the loop variable in terms
of a suitable basis of momenta (constructed from the
external momenta and arbitrarily chosen reference momenta),
the cut-conditions
impose kinematical bounds on the values of the components
of the loop momentum in this basis.
According to the number of cuts, some component (when not all),
are completely frozen, while others remain as free-variables\footnote{
The variables not frozen by the cut-conditions correspond to the integration variables 
of the phase-space integral}.
The integrand, evaluated at a value of
the loop momentum chosen among the solutions of a given cut, 
is a {\it polynomial} whose variables are the components of the loop
momentum not frozen by the cut-conditions:
the zero-th order coefficient of such a polynomial corresponds to the coefficient of
the master integral one is interested in;
the other terms of the polynomial, addressed to as {\it spurious} terms, 
do not contribute to the cut (since they vanish upon integration), but the are needed later on for
extracting the coefficients of lower-point master integrals.\\
To extract all the coefficients of this polynomial, one establishes 
a system, as said above, generated by evaluating numerically the integrand
for values of loop momentum chosen within the solutions of the cut-conditions (parametrized by its free-components).
The number of numerical evaluations must be the same as the number of 
the unknown coefficients to be determined.

The goal of the current paper is to exploit the polynomial structure of the integrand
and the freedom in choosing the solutions of the cuts, 
used as numerical points for the evaluation
of the integrand, to improve the system-solving algorithm.
By selecting the variables of each polynomial to be proportional to the roots of unity,
the extraction of the polynomial's coefficients is carried through {\it projections},
using the orthogonality relation among wave-planes,
rather than by system inversion. The basic principle underlying this procedure
is the same as for the Discrete Fourier Transform. 
The solutions accordingly obtained may help in getting a substantial
gain in computing-time.
The effective benefit of the new implementation is mostly experienced for the extraction of 
the coefficients of the 3- and 2-point functions, 
since in the 4- and 1-point cases the polynomial structure is very simple:
a degree-1 polynomial, in the case of the 4-point; an effective degree-0
polynomial, in the case of the 1-point functions.

The paper is organized as follows.
In Section 2, we recall the basic features of the OPP-reduction, by classifying the
polynomial structures generated by the the multiple-cut integrand.
In Section 3, we introduce the {\it projections} used for extracting the coefficients out of a polynomial,
which are explicitly applied, in Sections 4 and 5, for the extraction of 
the coefficients of the 3- and 2-point functions, respectively.
Finally, in Section 6, we apply the optimized reduction to the computation 
of the Next-to-Leading-Order QCD corrections to 
the scattering amplitude of the process $u \bar{d} \to W^+ W^- W^+$. 

\section{{\tt OPP}-Reduction}

The starting point of the {\tt OPP} reduction method~\cite{Ossola:2006us,Ossola:2007bb}
is the general expression for the {\it integrand} of a generic $m$-point 
one-loop (sub-)amplitude that, using dimensional
regularization, can be written as
\bqa
\label{eq:1}
A(\bar q)= \frac{N(q)}{\db{0}\db{1}\cdots \db{m-1}}\,,~~~
\db{i} = ({\bar q} + p_i)^2-m_i^2\,,~~~ p_0 \ne 0\,.
\eqa
We use a bar to denote objects living in $n=~4+\epsilon$  
dimensions; therefore we have  $\bar q^2= q^2+ \tld{q}^2$, where
$\tld{q}^2$ is $\epsilon$-dimensional and $(\tld{q} \cdot q) = 0$.
$N(q)$ is the $4$-dimensional part of the numerator of the amplitude. 
If needed, the $\epsilon$-dimensional part of the numerator should 
be treated separately, as explained in~\cite{Pittausimple,Ossola:2008xq}.
$N(q)$ depends on the $4$-dimensional denominators
$\d{i} = ({q} + p_i)^2-m_i^2$ as follows
\bqa
\label{eq:2}
N(q) &=&
\sum_{i_0 < i_1 < i_2 < i_3}^{m-1}
\left[
          d( i_0 i_1 i_2 i_3 ) +
     \tld{d}(q;i_0 i_1 i_2 i_3)
\right]
\prod_{i \ne i_0, i_1, i_2, i_3}^{m-1} \d{i} \nl
     &+&
\sum_{i_0 < i_1 < i_2 }^{m-1}
\left[
          c( i_0 i_1 i_2) +
     \tld{c}(q;i_0 i_1 i_2)
\right]
\prod_{i \ne i_0, i_1, i_2}^{m-1} \d{i} \nl
     &+&
\sum_{i_0 < i_1 }^{m-1}
\left[
          b(i_0 i_1) +
     \tld{b}(q;i_0 i_1)
\right]
\prod_{i \ne i_0, i_1}^{m-1} \d{i} \nl
     &+&
\sum_{i_0}^{m-1}
\left[
          a(i_0) +
     \tld{a}(q;i_0)
\right]
\prod_{i \ne i_0}^{m-1} \d{i} \nl
     &+& \tld{P}(q)
\prod_{i}^{m-1} \d{i}\,. \eqa
Inserted back in \eqn{eq:1}, this expression
simply states the multi-pole nature of any $m$-point one-loop amplitude,
that, clearly, contains a pole for any
propagator in the loop, thus one has terms ranging from 1 to $m$ poles.

The coefficients of the poles can be further split in two pieces.
A piece that still depend on $q$ (the terms
$\tld{d},\tld{c},\tld{b},\tld{a}$), that vanishes upon integration,
and a piece that do not depend on q (the terms $d,c,b,a$).
Such a separation is always possible, as shown in~\cite{Ossola:2006us}, 
and, with this choice, the latter set of coefficients is therefore immediately
interpretable as the ensemble of the 
coefficients of all possible 4, 3, 2, 1-point
one-loop functions contributing to the amplitude.
Notice that the term with no poles, namely that one proportional to
$\tld{P}(q)$ is polynomial and vanishes upon integration
in dimensional regularization. Moreover, it can be shown that in 
the renormalizable gauge $\tld{P}(q)= 0$, even before integration.

\subsection{Top-Down Polynomial Structures}

Since the scalar 1-, 2-, 3-, 4-point functions are known,
the problem of  computing 
the one-loop amplitude is simply reduced to the algebraical 
problem of fitting
the coefficients $d,c,b,a$ by evaluating the function $N(q)$
a sufficient number of times, at different values of $q$,
and then inverting the system.

This task can be achieved very efficiently by singling out
particular choices of $q$ such that, systematically,
4, 3, 2 or 1, among all possible denominators $\d{i}$ vanish.
In \cite{Ossola:2006us}, it was shown that by proceeding {\it top-down} from 
quadruple-cuts to single-cuts, it is possible to construct a particularly
simple system of equations that can be solved analytically, 
whose solutions yield the complete reconstruction of the unknown coefficients.

\begin{itemize}

\item{\bf Quadruple-cut. \ }
When $q$ is solution of \bqa \label{eq:8a} \d{0}= \d{1}= \d{2}=
\d{3} = 0\,. \eqa

\bqa
N(q) &=& [d(0123) + \tld{d}(q;0123)] \prod_{i\ne 0,1,2,3}\d{i} (q)\, \nl\nl
     &\equiv& R(q) \prod_{i \ne 0,1,2,3} \d{i}(q)
\eqa
where $R(q)$ has a polynomial structure with 2 terms.

\item{\bf Triple-cut. \ }
At this stage all $d$ and $\tld{d}$ coefficients are known.
When $q$ is solution of
\bqa
\label{eq:8b}
\d{0}= \d{1}= \d{2}= 0~~~{\rm and}~~~\d{i} \ne 0~~ \forall i \ne 0,1,2
\eqa
\eqn{eq:2} reads
\bqa
N(q) &-& \sum_{2 <i_3}[d(012i_3)
           + \tld{d}(q;012i_3)]\prod_{i \ne 0,1,2,i_3}\d{i}(q) \nl\nl
&\equiv& R^\prime(q) \prod_{i \ne 0,1,2} \d{i}(q)
= [c(012) + \tld{c}(q;012)]\prod_{i \ne 0,1,2} \d{i}(q)\,,
\label{eq:Rprime:def}
\eqa
where $R^{\prime}(q)$ has a polynomial structure with 7 terms.

\item{\bf Double-cut. \ }
At this stage all $d$, $\tld{d}$, $c$ and $\tld{c}$  coefficients are known.
When $q$ is solution of
\bqa
\label{eq:14}
\d{0}= \d{1}= 0~~~{\rm and}~~~\d{i} \ne 0~~ \forall i \ne 0,1\,
\eqa
\eqn{eq:2} reads
\bqa
 N(q)
&-& \sum_{1 < i_2 < i_3}[d(01i_2i_3)
+ \tld{d}(q;01i_2i_3)]  \prod_{i \ne 0, 1,i_2,i_3}\d{i} \nl \nl
&-& \sum_{1 < i_2}[c(01i_2)
+ \tld{c}(q;01i_2)] \prod_{i \ne 0, 1, i_2}\d{i} \nl \nl
&\equiv& {R}^{\prime\prime}(q) \prod_{i\ne 0,1}\d{i}(q)
 = [b(01) + \tld{b}(q;01)]\prod_{i\ne 0,1}\d{i}(q)\,,
\label{eq:Rsecond:def}
\eqa
where $R^{\prime\prime}(q)$ has a polynomial structure with 9 terms.

\item{\bf Single-cut. \ }
In massless theories all 1-point functions, namely all tadpoles,
vanish, also implying that, in such cases, one does not need to know
all the $\tld{b}$ coefficients.
However, in general, also the coefficients of the tadpoles are
required. Therefore we discuss how to extract them.

At this stage we assume to know all the
$d$, $\tld{d}$, $c$, $\tld{c}$, $b$ and $\tld{b}$ coefficients and,
when $q$ is solution of
\bqa
\label{eq:sola}
\d{0}= 0~~~{\rm and}~~~\d{i} \ne 0~~ \forall i \ne 0\,,
\eqa
\eqn{eq:2} reads
\bqa
 N(q)
&-& \sum_{0 <i_1 < i_2 < i_3}[d(0i_1i_2i_3)
+ \tld{d}(q;0i_1i_2i_3)]  \prod_{i \ne 0, i_1, i_2,i_3}\d{i} \nl \nl
&-& \sum_{0 <i_1 < i_2}[c(0i_1i_2)
+ \tld{c}(q;0i_1i_2)] \prod_{i \ne 0, i_1, i_2}\d{i} \nl \nl
&-& \sum_{0< i_1}[b(0i_1)
+ \tld{b}(q;0i_1)] \prod_{i \ne 0, i_1}\d{i} \nl \nl
&\equiv& {R}^{\prime\prime\prime}(q) \prod_{i\ne 0}\d{i}(q)
 = [a(0) + \tld{a}(q;0)]\prod_{i\ne 0}\d{i}(q)\,.
\eqa
where $R^{\prime\prime\prime}(q)$ has a polynomial structure with 5 terms.
We notice that the spurious coefficients $\tilde{a}_{i}(0) \ (i=1,...,4)$ are never needed,
because they would be necessary only to extract what we called $\tilde{P}(q)$, that,
as already observed, is irrelevant.
Therefore one focuses directly on the extraction of the tadpole-coefficient 
$a(0)$.

\end{itemize}

\bigskip

\bigskip

\noindent
We remark, that the polynomials we are going to deal with, {\it i.e.}
$R(q), R^{\prime}(q), R^{\prime\prime}(q)$, and $R^{\prime\prime\prime}(q)$, 
share a common structure: 
a {\it spurious} structure, depending on the loop variable, $q$, which does not 
contribute to the cut-integral;
and a single $q$-independent term, which corresponds to the actual coefficient of the 
master integral identified by the cuts\footnote{The only exception to this pattern will be the
structure of the double-cut coefficients, that also includes a 
$q$-independent non vanishing term, in order to avoid numerical instabilities~\cite{Ossola:2007bb}.}.

\section{Polynomial Structures and Discrete Fourier Transform}

After the general structure of \eqn{eq:2} is established, as we illustrated in the previous section,
the calculation of the scattering amplitude reduces to the problem of extracting the coefficients of multivariable polynomials, generated at every step of the multiple-cut analysis.

Let us show how it is possible to extract efficiently the coefficients
of a polynomial of degree $n$ in the variable $x$, say $P_n(x)$, defined as,
\bea
P_n(x) = \sum_{\ell=0}^n \ c_\ell \ x^\ell \ ,
\label{eq:GeneralPn}
\eea
by means of {\it projections}, according to the same principle underlying
the Discrete Fourier Transform which works as follows.

\subsection{Discrete Fourier Transform}

Consider the function $F$, known only numerically in $N$ points, $F_n \
(n=0,...,N-1)$.
Each of this values, admits a DFT, defined as,
\bea
F_n \equiv \sum_{k=0}^{N-1} f_k \  e^{-2 \pi i { k\over N} \ n} \ .
\eea
The coefficients $f_k$, can be found by using the orthogonality relation,
\bea
\sum_{n=0}^{N-1} e^{2 \pi i {k \over N} \ n} \ e^{-2 \pi i{k^\prime \over N} \ n}
 = N \ \delta_{k k'} \ .
\eea
with the result
\bea
f_k = {1 \over N}
      \sum_{n=0}^{N-1}
      F_n \ 
      e^{2 \pi i {k \over N} \ n} \ .
\eea

\subsection{Projections}

Let us see how the above procedure can
be implemented for
extracting the coefficients $c_\ell$'s of the
polynomial in Eq.(\ref{eq:GeneralPn}) by {\it projections}. 
\begin{enumerate}
\item
Generate the set of discrete values $g_k \ (k=0,...,n)$,
\bea
g_k = P_n(x_k) = 
\sum_{\ell=0}^n \ c_\ell \ \rho^\ell \ e^{-2 \pi i {k \over (n+1)} \ell} \ ,
\eea
by evaluating $P_n(x)$ at the points
\bea
 x_k = \rho \ e^{-2 \pi i {k \over (n+1)}} \ .
\eea

\item
Using the orthogonality relations for the wave planes,
one can obtain the coefficient $c_\ell$ simply by,
\bea
c_\ell &=& {\rho^{-\ell} \over n+1}
         \sum_{k=0}^n \ g_k \  e^{2 \pi i {k \over (n+1)} \ell}
\eea
\end{enumerate}

\bigskip

\bigskip

This procedure is very general and can be applied as long as 
one needs to know the coefficients of any polynomial.
In fact, it has been recently suggested in \cite{Britto:2008vq},
for computing the coefficients of polynomials in the $(-2 \epsilon)$-dimensional mass
parameter. \\
The projections could be extended also to the case of multi-variables polynomials,
along the same line of the multi-dimensional DFT.
Since we aim to minimize the computational load, we keep
the number of numerical evaluations of each polynomial to be
the same as the number of its coefficients.
To this aim, we will see that 
the coefficients of a multi-variable polynomial
can be equivalently found by breaking it in several one-variable polynomials, 
obtained from the former by freezing the values of the other variables,
yielding still the use of the (one-dimensional) projections described above.

The next two sections will be devoted to the application of
the projection procedure for the
extraction of the coefficients of the 3- and 2-point functions,
respectively out of $R^{\prime}(q)$, and $R^{\prime\prime}(q)$.
We won't discuss hereby the reconstruction of the coefficients of the 4- and 1-point functions, 
because the polynomial structures of $R(q)$ and $R^{\prime\prime\prime}(q)$ is very simple, and
the result of the projection procedure would be the same as the one given in~\cite{Ossola:2006us}.

\section{The coefficient of the 3-point functions}

In this section,
we show how to apply the properties of
orthogonal functions for extracting the
coefficient of the 3-point functions. \\
Let's begin from Eq.(2.21) of \cite{Ossola:2006us},
\bea
R^\prime(q) = c(012) + 
\sum_{j=1}^3 
\left\{
  \tilde{c}_{1j}(012)[(q + p_0) \cdot \ell_3 ]^j
+ \tilde{c}_{2j}(012)[(q + p_0) \cdot \ell_4 ]^j
\right\}
\eea
where $R^\prime(q)$ appeared in Eq.(\ref{eq:Rprime:def}).
By substituting the parametrization of $q$,
solution of the triple-cut conditions
given in \cite{Ossola:2006us}, and recalled in App. \ref{app:3ple},
\bea
q = -p_0 + x_1 \ell_1 + x_2 \ell_2 + x_3 \ell_3 + x_4 \ell_4
\label{eq:loop:mom:3ple}
\eea
one obtains
\bea
R^\prime(q) = c(012) + 
\sum_{j=1}^3 
\left\{
  \tilde{c}_{1j}(012) \ (\ell_3 \cdot \ell_4)^j \ x_4^j
+ \tilde{c}_{2j}(012) \ (\ell_3 \cdot \ell_4)^j \ x_3^j 
\right\} \ ,
\eea
This expression can be read as a polynomial in $x_3$ and $x_4$,
whose canonical form reads,
\bea
P(x_3,x_4) &=&
  c_{0,0}
+ c_{1,0} \ x_3
+ c_{2,0} \ x_3^2
+ c_{3,0} \ x_3^3
+ c_{0,1} \ x_3
+ c_{0,2} \ x_4^2
+ c_{0,3} \ x_4^3
\label{OPP:Poly:x3x4}
\eea 
with the following relations among the coefficients
\bea
c_{0,0} &=& c(012) \ , \nonumber \\
{c_{j,0} } &=&  c_{2j}(012) \ (\ell_3\cdot\ell_4)^j \nonumber \\
{c_{0,j} } &=& c_{1j}(012) \ (\ell_3\cdot\ell_4)^j
\label{TRIPLE:fromME2OPP}
\eea
Therefore computing the {\tt OPP}-coefficients is 
equivalent to the computation
of the 7 coefficients of Eq.(\ref{OPP:Poly:x3x4}).

\subsection{Projections}

\noindent
The extraction of the coefficients $c_{i,j}$ is performed, in the framework
   of the original {\tt OPP}-method and in {\tt CutTools}, by choosing a redundand
   set of solutions in order to avoid some fake singularities occurring
   in kinematical points in which $C=0$. In practice, this is obtained by
   doubling the number of calls to the numerator function $N(q)$, by roughly
   doubling the computation time. \\
   The same problem can be more efficiently solved with the help
   of the proposed {\it projection} method. By using it, it is in fact very easy
   to find a set of solutions for which the point $C=0$ is never singular and
   for which $N(q)$ is called just as many time as the number of needed
   coefficients. This is done at the price of creating a new fake
   singularity at $C=1$, but, when this situation occurr, one can use
   the original solution. The described procedure explicitly shows the
   flexibility of the proposed {\it projection} method. \\

\noindent
To extract the 7 coefficients $c_{ij}$ by projections,
we take 7 values of $P$, grouped in 2 sets $(4+3)$,
\bea
g_{1,k} &=& P(x_{3k}, x_{4k}) \ , 
\qquad 
x_{3k} = C \ e^{- 2 \pi i {k \over 4}} \ , \ 
x_{4k} =     e^{  2 \pi i {k \over 4}} \ , 
\quad (k=0,1,2,3) \nonumber \\
g_{2,k} &=& P(x_{3k}, x_{4k}) \ , 
\qquad 
x_{3k} =     e^{- 2 \pi i {k \over 3}} \ , \ 
x_{4k} = C \ e^{  2 \pi i {k \over 3}} \ , 
\quad (k=0,1,2) 
\eea
where $C$ is given in App. \ref{app:3ple}.
\noindent
Then, we construct the auxiliary functions
\bea
\mu(1,m,n) = {1 \over 4} \sum_{k=0}^3 \ g_{1,k} \ 
e^{2 \pi i {k \over 4} (m-n)} \ 
\eea
with:
$(m,n) = (0,0), (0,1), (0,2), (1,0)$.
\bea
\mu(2,m,n) = {1 \over 3} \sum_{k=0}^2 \ g_{2,k} \ 
e^{2 \pi i {k \over 3} (m-n)} \ 
\eea
with:
$(m,n) = (0,0), (0,1), (1,0)$.

\noindent
In terms of these auxiliary functions, the coefficients read,
\bea
c_{0,0}&=& \mu (1,0,0) \\
c_{1,0}&=& -\frac{1}{C^{12}-1} \Bigg(-\mu (1,1,0) C^{11}-\mu (1,0,0) C^8+\mu (2,0,0) C^8-\mu
   (1,0,1) C^5+\mu (2,0,1) C^4\nonumber \\ && \qquad \qquad-\mu (1,0,2) C^2+\mu (2,1,0)
\Bigg)\\
c_{2,0}&=& -\frac{1}{C^{12}-1} \Bigg(-\mu (1,0,2)
   C^{10}+\mu (2,1,0) C^8-\mu (1,1,0) C^7-\mu (1,0,0) C^4\nonumber \\ && \qquad \qquad
+\mu (2,0,0) C^4-\mu (1,0,1) C+\mu
   (2,0,1)
\Bigg)\\
c_{3,0}&=& -\frac{1}{C^{12}-1}
\Bigg(-\mu (1,0,1) C^9+\mu (2,0,1) C^8-\mu (1,0,2) C^6 
+\mu (2,1,0) C^4\nonumber \\ && \qquad \qquad -\mu (1,1,0) C^3-\mu
   (1,0,0)+\mu (2,0,0) \Bigg) \\
c_{0,1}&=& -\frac{1}{C^{12}-1} \Bigg(-\mu (2,0,1) C^{11}+\mu (1,0,2) C^9-\mu
   (2,1,0) C^7 
+\mu (1,1,0) C^6+\mu (1,0,0) C^3\nonumber \\ && \qquad \qquad
-\mu (2,0,0) C^3+\mu (1,0,1)
\Bigg)\\
c_{0,2}&=& -\frac{1}{C^{12}-1} \Bigg( -\mu (2,1,0) C^{10}+\mu (1,1,0) C^9+\mu (1,0,0) C^6-\mu (2,0,0) C^6+\mu (1,0,1)
   C^3\nonumber \\ && \qquad \qquad-\mu (2,0,1) C^2+\mu (1,0,2)
\Bigg)\\
c_{0,3}&=& -\frac{1}{C^{12}-1} \Bigg(\mu (1,0,0) C^9-\mu (2,0,0)
   C^9+\mu (1,0,1) C^6-\mu (2,0,1) C^5+\mu (1,0,2) C^3\nonumber \\ && \qquad \qquad-\mu (2,1,0) C+\mu (1,1,0)
\Bigg)
\eea

\noindent
Finally, to obtain the {\tt OPP}-coefficients, 
simply use Eqs.(\ref{TRIPLE:fromME2OPP}). \\

\section{The coefficient of the 2-point functions}

In this section,
we show how to apply the properties of
orthogonal functions for extracting the
coefficient of the 2-point functions. \\
Let's begin from Eq.(B.7) of \cite{Ossola:2007bb}
\bea
R^{\prime\prime}(q) &=& 
  b
+ {\hat b}_0 [(q+p_0) \cdot v] 
+ {\hat b}_{00} [(q+p_0) \cdot v]^2 \nonumber \\ &&
+ {\tilde b}_{11} [(q+p_0) \cdot \ell_7] 
+ {\tilde b}_{21} [(q+p_0) \cdot \ell_8] \nonumber \\ &&
+ {\tilde b}_{12} [(q+p_0) \cdot \ell_7]^2 
+ {\tilde b}_{22} [(q+p_0) \cdot \ell_8]^2 \nonumber \\ &&
+ {\tilde b}_{01} [(q+p_0) \cdot \ell_7] [(q+p_0) \cdot v] 
+ {\tilde b}_{02} [(q+p_0) \cdot \ell_8] [(q+p_0) \cdot v] 
\eea
where $R^{\prime\prime}(q)$ appeared in Eq.(\ref{eq:Rsecond:def}).
By substituting the parametrization of $q$,
the solution of the double-cut conditions
given in Eq.(B.6) of \cite{Ossola:2007bb}, and recalled in App. \ref{app:2ple},
\bea
q = -p_0 + y k_1 + y_v v + y_7 \ell_7 + x_8 \ell_8
\eea
one obtains
\bea
N(q) &=&
  b
+ {\hat b}_0 [y k_1 \cdot v] 
+ {\hat b}_{00} [y k_1 \cdot v]^2 \nonumber \\ &&
+ {\tilde b}_{11} [y_8  \ell_8 \cdot \ell_7] 
+ {\tilde b}_{21} [y_7  \ell_7 \cdot \ell_8] \nonumber \\ &&
+ {\tilde b}_{12} [y_8  \ell_8 \cdot \ell_7]^2 
+ {\tilde b}_{22} [y_7  \ell_7 \cdot \ell_8]^2 \nonumber \\ &&
+ {\tilde b}_{01} [y_8  \ell_8 \cdot \ell_7] [y k_1 \cdot v]  
+ {\tilde b}_{02} [y_7  \ell_7 \cdot \ell_8] [y k_1 \cdot v] 
\eea
This expression can be read as a polynomial in $y$, $y_7$ and $y_8$,
whose canonical form reads,
\bea
P(y, y_7, y_8) &=&
  a_{000}
+ a_{100} \ y
+ a_{200} \ y^2 \nonumber \\ &&
+ a_{010} \ y_7
+ a_{020} \ y_7^2
+ a_{001} \ y_8
+ a_{002} \ y_8^2 \nonumber \\ &&
+ a_{110} \ y \ y_7
+ a_{101} \ y \ y_8 \ ,
\label{OPP:Poly:yy7y8}
\eea 
with the following relations among the coefficients
\bea
a_{000} &=& b \nonumber \\
a_{100} &=& {\hat b}_{0} (k_1 \cdot v) \nonumber \\
a_{200} &=& {\hat b}_{00} (k_1 \cdot v)^2 \nonumber \\
a_{010} &=& {\tilde b}_{21} (\ell_7 \cdot \ell_8) \nonumber \\
a_{020} &=& {\tilde b}_{22} (\ell_7 \cdot \ell_8)^2 \nonumber \\
a_{001} &=& {\tilde b}_{11} (\ell_7 \cdot \ell_8) \nonumber \\
a_{002} &=& {\tilde b}_{12} (\ell_7 \cdot \ell_8)^2 \nonumber \\
a_{110} &=& {\tilde b}_{02} (\ell_7 \cdot \ell_8) (k_1 \cdot v) \nonumber \\
a_{101} &=& {\tilde b}_{01} (\ell_7 \cdot \ell_8) (k_1 \cdot v)
\label{DOUBLE:fromME2OPP}
\eea

\subsection{Projections}

\noindent
Here we explicitly illustrate a solution that avoids the problem
    of doubling the number of calls to the numerator function $N(q)$
    (used in the original implementation of the {\tt OPP}-method)
    due to the appearence of fake singularities when $F_0 = 0$.\\

\noindent
To extract the 9 coefficients $a_{ijk}$ by projections,
we take 9 values of $P$, grouped in 5 sets $(3+2+2+1+1)$,
\bea
g_{00h} &=& P(0, y_{7h}, y_{8h}) \ , 
\qquad 
y=0, \ 
y_{7h} = F_0 \ e^{- 2 \pi i {h \over 3}}  \ , \ 
y_{8h} =       e^{  2 \pi i {h \over 3}} \ , 
\quad (h=0,1,2) \nonumber \\
&& \nonumber \\
g_{0h0} &=& P(0, y_{7h}, y_{8h}) \ , 
\qquad 
y=0 \ , \ 
y_{7h} =       e^{- 2 \pi i {h \over 2}} \ , \
y_{8h} = F_0 \ e^{  2 \pi i {h \over 2}} 
\quad (h=0,1) \nonumber \\
&& \nonumber \\
g_{-10h} &=& P(-1, y_{7h}, y_{8h}) \ , 
\qquad 
y=-1 \ , \ 
y_{7h} = F_{-1} \ e^{- 2 \pi i {h \over 2}}  \ , \ 
y_{8h} =          e^{  2 \pi i {h \over 2}}
\quad (h=0,1) \nonumber \\
&& \nonumber \\
g_{-1} &=& P(-1, 1, F_{-1}) \ , 
\qquad 
y=-1 \ , \ 
y_{7} = 1 \ , \ 
y_{8} = F_{-1}  
\nonumber \\
&& \nonumber \\
g_{1} &=& P(1, F_{1},1) \ , 
\qquad 
y= 1 \ , \ 
y_{7} = F_{1} \ , \ 
y_{8} =  1 \nonumber \\
\eea
where the definition of $F_{y}$ is given in App. \ref{app:2ple}.

\noindent
Then, we construct the auxiliary functions
\bea
\mu(1,0,m,n) =  
             {1 \over 3} \sum_{h=0}^2 \ g_{00h} \ 
e^{ 2 \pi i {h \over 3} (m-n)} \ ,
\qquad (m,n) = (0,0), (0,1), (1,0)
\eea

\bea
\mu(2,0,m,n) = 
             {1 \over 2} \sum_{h=0}^1 \ g_{0h0} \ 
e^{ 2 \pi i {h \over 2} (m-n)} \ ,
\qquad (m,n) = (0,0), (1,0)
\eea

\bea
\mu(1,-1,m,n) =  
             {1 \over 2} \sum_{h=0}^1 \ g_{-10h} \ 
e^{ 2 \pi i {h \over 2} (m-n)} \ ,
\qquad (m,n) = (0,0), (0,1)
\eea

\bea
\mu(2,-1,0,0) = g_{-1} \ 
\eea

\bea
\mu(1,1,0,0) = \ g_{1}
\eea
The coefficients can be expressed as linear combinations
of these auxiliary functions.
Since the expressions for generic
values of $F_{0}$, $F_{-1}$, $F_{1}$ are rather long,
we present the one obtained when $F_{0} = F_{-1} = 0$, which happens
when the propagators are massless, as in the case of
the application later discussed. Hence, the coefficient read,

\bea
a_{000}&=& \mu (1,0,0,0) \\ 
a_{100}&=& \frac{1}{2} \Bigg(\mu (1,0,0,0) F_1^2-\mu (2,0,0,0) F_1^2-\mu
   (1,-1,0,0) F_1+\mu (1,0,0,0) F_1 \nonumber \\ && \qquad +\mu (1,0,1,0) F_1+\mu (2,-1,0,0) F_1-\mu (2,0,0,0) F_1-2\mu (2,0,1,0) F_1 \nonumber \\ && \qquad-\mu (1,-1,0,0)+\mu (1,-1,0,1)-2 \mu (1,0,0,1)+\mu (1,1,0,0)\Bigg), \\ 
a_{200}&=& \frac{1}{2}
   \Bigg(\mu (1,0,0,0) F_1^2-\mu (2,0,0,0) F_1^2-\mu (1,-1,0,0) F_1+\mu (1,0,0,0) F_1
\nonumber \\ && \qquad+\mu (1,0,1,0)
   F_1+\mu (2,-1,0,0) F_1-\mu (2,0,0,0) F_1-2 \mu (2,0,1,0) F_1\nonumber \\ && \qquad
+\mu (1,-1,0,0)
+\mu (1,-1,0,1)-2 \mu(1,0,0,0)\nonumber \\ && \qquad -2 \mu (1,0,0,1)-2 \mu (1,0,1,0)+\mu (1,1,0,0)\Bigg), \\ 
a_{010}&=& \mu(2,0,1,0), \\ 
a_{020}&=& \mu (2,0,0,0)-\mu (1,0,0,0), \\
a_{001}&=& \mu (1,0,0,1), \\ 
a_{002}&=& \mu (1,0,1,0), \\ 
a_{110}&=& \mu (1,-1,0,0)-\mu (1,0,0,0)-\mu (1,0,1,0)-\mu (2,-1,0,0)+\mu (2,0,0,0)+\mu (2,0,1,0), \qquad \\ 
a_{101}&=& \mu(1,0,0,1)-\mu (1,-1,0,1) \ .
\eea

\noindent
Finally, to obtain the {\tt OPP}-coefficients, 
simply use Eqs.(\ref{DOUBLE:fromME2OPP})

\section{An example: QCD virtual corrections to $u \bar{d} \to W^+ W^- W^+$ at NLO}

As an example of application of the optimized algorithm, we present the results for the 
pentagon and box diagrams contributing to the scattering 
amplitudes of $u \bar{d} \to W^+ W^- W^+$ at NLO in QCD. The complete cross section for
the production of $W^+ W^- W^+$ at LHC, as well as $W^+ W^- Z$, $W^+ Z Z$ and $Z Z Z$ is presented in \cite{3bosons}.

The main purpose of this application is to test the improvements on the {\tt OPP}-reduction algorithm, both in terms of stability and efficiency, after the changes proposed in this paper have been included.

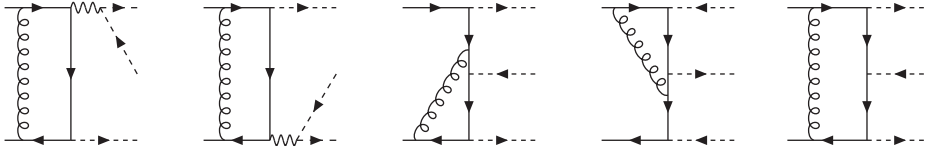
\begin{figure}[ht]
\begin{center}
\begin{picture}(325,50)(-50,0)
        \ArrowLine(-75,50)(-50,50)
        \ArrowLine(-50,0)(-75,0)
        \ArrowLine(-50,50)(-50,0)
	\Photon(-50,50)(-39,50){2}{3}
        \DashArrowLine(-39,50)(-25,50){2}
        \DashArrowLine(-50,0)(-25,0){2}
        \DashArrowLine(-25,25)(-39,50){2}
        \Gluon(-68,0)(-68,50){2}{8}
        \ArrowLine(0,50)(25,50)
        \ArrowLine(25,0)(0,0)
  	\ArrowLine(25,50)(25,0)
	\Photon(25,0)(35,0){2}{3}
        \DashArrowLine(35,0)(50,0){2}
        \DashArrowLine(50,25)(35,0){2}
        \DashArrowLine(25,50)(50,50){2}
        \Gluon(8,0)(8,50){2}{8}
        \ArrowLine(75,50)(100,50)
        \ArrowLine(100,0)(75,0)
        \ArrowLine(100,25)(100,0)
	\ArrowLine(100,50)(100,25)
        \DashArrowLine(100,50)(125,50){2}
        \DashArrowLine(100,0)(125,0){2}
        \DashArrowLine(125,25)(100,25){2}
        \Gluon(100,34)(80,0){-2}{6}
        \ArrowLine(150,50)(175,50)
        \ArrowLine(175,0)(150,0)
        \ArrowLine(175,25)(175,0)
	\ArrowLine(175,50)(175,25)
        \DashArrowLine(200,50)(175,50){2}
        \DashArrowLine(175,25)(200,25){2}
        \DashArrowLine(200,0)(175,0){2}
        \Gluon(175,17)(155,50){2}{6}
        \ArrowLine(220,50)(250,50)
        \ArrowLine(250,0)(220,0)
        \ArrowLine(250,25)(250,0)
	\ArrowLine(250,50)(250,25)
        \DashArrowLine(250,50)(275,50){2}
        \DashArrowLine(250,0)(275,0){2}
        \DashArrowLine(275,25)(250,25){2}
        \Gluon(230,0)(230,50){2}{8}
      \end{picture}
     \end{center} \caption{Box and pentagon diagrams contributing to virtual QCD corrections
to $u \bar{d} \to W^+ W^- W^+$. Wavy lines can be either Z or photons, dashed lines represent $W^+$ and $W^-$.
Diagrams involving an exchange in the two final $W^+$ should also be considered.} \label{topo}
\end{figure}

The complete calculation of NLO QCD virtual correction to $u \bar{d} \to W^+ W^- W^+$, neglecting the contributions that depend on the Higgs boson, involves the reduction of 53 diagrams. 
The topologies of pentagon and box diagrams contributing to this process are depicted in Fig.~\ref{topo}. Overall, we have 2 pentagons and 12 boxes.

Most of the computing time in the calculation is spent in the evaluation of the coefficients for 3-point
and 2-point scalar function arising from the reduction of the diagrams illustrated.
For example, the reduction of the each pentagon implies the evaluation of ten sets of c-type coefficients
and ten sets of b-type coefficients, involving each seven and nine coefficients respectively.
It is therefore very important to have efficient routines to achieve this task. As a comparison, the evaluation
of the d-type coefficients is only performed five times for each pentagon and involves only two coefficients.

For the purpose of this test, we fix the external momenta to the specific phase-space configuration of \eqn{momwww} and the polarization vectors for the vector bosons. 
To perform the calculation, we use the {\tt OPP}-reduction method, with and without the optimization technique
illustrated before. The coefficients determined in this manner should be multiplied by the corresponding scalar 
integrals. Since, in the process that we are studying, no $q$-dependent massive propagator appear, we will only need massless scalar integrals. They are computed using the package {\tt OneLOop} written by A.~van~Hameren~\cite{avh}. Finally, we sum the contributions coming from the various pentagons and boxes: 
the results obtained are presented in Table~\ref{tabella}.
\bea
p_1 & = & \{ 500, 0, 500, 0 \} \nonumber \\ 
p_2 & = &\{ 500, 0, -500, 0 \} \nonumber \\  
 p_3 & = & \{ 276.212, 97.7237, -56.2856, 238.9792 \} \nonumber \\ 
 p_4 & = &\{ 486.8926, 213.4030, 37.7214, -428.5282 \} \nonumber \\ 
p_5 & = & p_1 + p_2 - p_3 - p_4 \label{momwww}
\eea
The results obtained with the two different implementations of the 
algorithm are, of course, in perfect agreement. However, with the optimized 
version we can roughly improve the efficiency of factor 2.
As a first test, we checked the improvements on the timing of the two
new routines alone, separating them from the rest of the reduction.
We experience a reduction in the computing time of about 60\% for the 
system of the c-coefficients, and about 50\% for the system of 
the b-coefficient.

In the overall evaluation of the amplitudes, the optimized routines 
are combined with other parts of the program that remain unchanged.
This involves, for example, the initialization, the evaluation of 
scalar integrals, the evaluation of d- and a-coefficients
and rational parts, and the summing over all contributions. We still 
retain, however, an overall improvement in the computing time of about 40\%.

Concerning the stability issues, we tested the new routines for a wide set of 
phase-space points. We do not observe significant improvements respect to 
the previous implementation.

\begin{table}[ht] \label{tabella}
\begin{center}
 \begin{tabular}{|c|c|}
   \hline
  Polarization & $|A|$ \\ 
   \hline
  $0$ $0$ $0$  & 28.2435 \\
  $0$ $0$ $-$  & 5.13851  \\
  $0$ $0$ $+$ & 10.7870   \\
  $0$ $-$ $0$  & 1.923741 \\
  $0$ $-$ $+$  & 0.718415 \\
  $0$ $+$ $0$ & 7.43599 \\ 
  $0$ $+$ $-$  & 1.95506 \\
  $-$ $+$ $-$  & 0.276058 \\
  $-$ $+$ $+$  & 0.402302 \\
  $+$ $+$ $+$  & 0.875457 \\
  \hline 
 \end{tabular} \caption{Absolute value of the amplitudes
for different configurations of the polarization of the vector bosons
(accounting for 5-point and 4-point Feynman diagrams only).
The results are expressed in unit of $e^3$.}
\end{center}
\end{table}

\section{Conclusions}

We presented an optimization of the
reduction algorithm of one-loop amplitudes in terms of
master integrals. That is based on the exploitation
of the polynomial structure of the integrand
when evaluated at values of the loop-momentum
fulfilling multiple cut-conditions, as
emerged in the {\tt OPP}-method.
Accordingly, the integrand, evaluated at a value of
the loop momentum chosen among the solutions of a given cut, 
is a {\it polynomial} whose variables are the components of the loop
momentum not frozen by the cut-conditions:
the zero-th order coefficient of such a polynomial corresponds to the coefficient of
the master integral one is interested in;
the other terms of the polynomial, though not contributing 
to the cut, are needed for the later 
determination of the coefficients of lower-point master integrals. 

To extract all the polynomial coefficients, one establishes 
a system generated by evaluating numerically the integrand
for values of loop variable chosen within the solutions of the cut-conditions 
(parametrized by its free-components).
The freedom in choosing the solutions of the cuts
has been hereby exploited to improve the system-solving algorithm.
By selecting the variables of each polynomial to be proportional to the {\it primitive 
roots of unity},
the extraction of the polynomial's coefficients is carried through {\it projections},
using the same orthogonality relation underlying the Discrete Fourier Transform. 
The number of numerical evaluations is kept as low as the number of 
the unknown coefficients to be determined, by using one-dimensional projections
also in case of polynomials in more than one-variable.

The novel implementation was applied to the reduction of
the 4- and 5-point one-loop Feynman diagrams contributing 
the NLO QCD corrections to $u \bar{d} \to W^+ W^- W^+$, where we
experienced a reduction of the computational load.

\bigskip

The flexibility of the projection-procedure hereby presented
extends its range of applicability to tackle 
the determination of the coefficients of polynomial structures
wherever should this issue occur.
Moreover, we finally remark that the parametrization of the free (integration) variables 
as complex unitary phases yields as well a very effective performance of Cauchy's
residue theorem within the contexts of factorization- and unitarity-based methods,
where the on-shellness properties are naturally captured by polar structures 
in complex phases.

\bigskip

\bigskip

\noindent {\bf Acknowledgments} 

\noindent
C.G.P.'s and R.P.'s research was partially supported by the RTN
  European Programme MRTN-CT-2006-035505 (HEPTOOLS, Tools and Precision
  Calculations for Physics Discoveries at Colliders).
  The research of R.P. was also supported by MIUR under contract
  2006020509\_004 and by the MEC project FPA2006-05294. \\
G.O. and R.P. acknowledge the financial support of the ToK Program ``ALGOTOOLS'' (MTKD-CT-2004-014319). \\
P.M. thanks Ettore Remiddi and Thobias Motz for interesting discussions
on the use of orthogonal functions, and Joe Apostolico for special support.

\appendix

\section{The basis for the 3-point functions}
\label{app:3ple}

The loop momentum solution of the triple-cut, given
in Eq.(\ref{eq:loop:mom:3ple}), is expressed in terms
of auxiliary vectors defined as follows. \\
$\ell_1$ and $\ell_2$  are massless 4-vector satisfying
the relations
\bqa \label{eq:4b} k_1   = \ell_1 + \alpha_1
\ell_2\,,~~~k_2   = \ell_2 + \alpha_2 \ell_1 \,, \eqa with \bqa
\label{eq:4bb}
 k_i = p_i-p_0\,.
\eqa
Furthermore, in spinorial notation,
\bqa
\label{eq:4c}
\ell_3^\mu = \langle\ell_1| \gamma^\mu | \ell_2]\,,~~
\ell_4^\mu = \langle\ell_2| \gamma^\mu | \ell_1]\,
~~{\rm with}~~ (\ell_3 \cdot \ell_4) = -4(\ell_1 \cdot \ell_2)\,.
\eqa
The solution to \eqn{eq:4b} reads
\bqa
\label{eq:4d}
\ell_1 &=& \beta(k_1- \alpha_1 k_2)\,,~~~ \ell_2
~=~ \beta(k_2- \alpha_2 k_1)\,,\nl
\beta  &=& 1/(1- \alpha_1 \alpha_2)\,,~~~ \alpha_i~=~\frac{k_i^2}{\gamma}\,,\nl
\gamma &\equiv& 2 (\ell_1 \cdot \ell_2) = (k_1 \cdot k_2) \pm \sqrt{\Delta}
\,,~~~\Delta ~=~  (k_1 \cdot k_2)^2-k_1^2 k_2^2\,.
\eqa
we decompose
$q^\mu+ p_0^\mu$ in the basis formed by $\ell_1, \ell_2, \ell_3$, and $\ell_4$,
\bea
q = -p_0 + x_1 \ell_1 + x_2 \ell_2 + x_3 \ell_3 + x_4 \ell_4 \ ,
\eea
which is solution of the triple-cut,
\bqa
D_0 = D_1 = D_2 = 0\,.
\eqa
Due to the above constrains,
the coefficients of the loop decomposition
must fulfill the following relations,
\bqa
\label{eq:sold1}
&&x_1 = \frac{\beta}{\gamma}
[d_2 - \alpha_2 d_1 -d_0 (1- \alpha_2)]\,, \\
&&x_2 = \frac{\beta}{\gamma}
[d_1 - \alpha_1 d_2 -d_0 (1- \alpha_1)]\,, \\
&&x_3 \ x_4 = C \,,
\eqa
where 
\bqa
\label{eq:sold2}
&&C = \frac{1}{4} \left(x_1 x_2-\frac{d_0}{\gamma}
                  \right)\,, \\
&&d_i \equiv m_i^2 - k_i^2 \ .
\eqa

\section{The basis for the 2-point functions}
\label{app:2ple}

First, we introduce a massless arbitrary 4-vector $v$,
such that $ (v \cdot k_1) \ne 0$, that we use to rewrite
$k_1$ in terms of two massless 4-vectors (we also take  $\ell^2=0$)
\bqa
k_1= \ell + \alpha \,v\,,
\eqa
giving
\bqa
\gamma \equiv 2\,(k_1 \cdot v) = 2\,(\ell \cdot v)~~{\rm and}~~
\alpha= \frac{k_1^2}{\gamma}\,.
\eqa
Then, we introduce two additional independent massless  4-vectors
$\ell_{7,8}$ defined as
\bqa
\ell_7^\mu &=& \langle \ell| \gamma^\mu |v]\,,~~
\ell_8^\mu = \langle v| \gamma^\mu |\ell]\,,
\eqa
for which one finds
\bqa
(\ell_7 \cdot \ell_8) =  -2 \gamma\,,
\eqa
and we decompose
$q^\mu+ p_0^\mu$ in the basis formed by $k_1$, $v$, $\ell_7$ and $\ell_8$
\bqa
\label{eq:qexp}
q^\mu = -p_0^\mu+ y k_1^\mu+ y_v v^\mu + y_7 \ell_7^\mu + y_8 \ell_8^\mu\,,
\eqa
that fulfill the double-cut requirement
\bqa
D_0 = D_1 = 0\,.
\eqa
For a $q$ written as in \eqn{eq:qexp} this implies the system
\bqa
&&y_7 y_8 = F_y \nl
&&y_v     = \frac{d_1-d_0 - 2 y k_1^2}{\gamma}\,,
\eqa
where
\bqa
F_y = - \frac{1}{4 \gamma}
\left(
m_0^2-y\,(d_1-d_0)+y^2 k_1^2
\right)\,.
\eqa
We remark that when $m_0 = 0$, $F_{y}$ vanishes for $y=-1,0$.


\end{document}